\documentclass[twocolumn, superscriptaddress]{revtex4}

\pdfoutput=1

\usepackage{amsmath}
\usepackage{amssymb}
\usepackage{graphicx}
\usepackage{units}

\DeclareMathOperator{\erfc}{erfc}

\newcommand{\MI}{I}
\newcommand{\Nl}{{N_{\rm l}}}
\newcommand{\Nr}{{N_{\rm r}}}
\newcommand{\Sbin}{\hat S}

\usepackage{amsmath}
\usepackage{xcolor}
\usepackage{dsfont} 
\usepackage{xspace}

\newcommand{\ie}{\mbox{i.\hspace{0.125em}e.}\@\xspace}
\newcommand{\eg}{\mbox{e.\hspace{0.125em}g.}\@\xspace}

\newcommand{\kb}{k_{\rm B}}

\definecolor{plot1}{RGB}{6,115,183}
\definecolor{plot2}{RGB}{255,118,0}
\definecolor{plot3}{RGB}{0,169,25}
\definecolor{plot4}{RGB}{230,0,28}
\definecolor{plot5}{RGB}{0,0,0}
\definecolor{plotBlue}{RGB}{6,115,183}
\definecolor{plotOrange}{RGB}{255,118,0}
\definecolor{plotGreen}{RGB}{0,169,25}
\definecolor{plotRed}{RGB}{230,0,28}

\newcommand{\Eqref}[1]{\mbox{Eq.\hspace{0.25em}\eqref{#1}}}
\newcommand{\Eqsref}[1]{\mbox{Eqs.\hspace{0.25em}\eqref{#1}}}
\newcommand{\figref}[1]{\mbox{Fig.\hspace{0.25em}\ref{#1}}}

\newcommand{\diff}{\text{d}}

\newcommand{\difffrac}[2]{\frac{\diff #1}{\diff #2}}

\newcommand{\set}[1]{\{{#1}\}}

\DeclareMathOperator{\var}{var}
\DeclareMathOperator{\cov}{cov}

\newcommand{\mean}[1]{\langle #1 \rangle}

\DeclareFontFamily{U}{mathx}{\hyphenchar\font45}
\DeclareFontShape{U}{mathx}{m}{n}{<-> mathx10}{}
\DeclareSymbolFont{mathx}{U}{mathx}{m}{n}
\DeclareMathAccent{\widebar}{0}{mathx}{"73}
\newcommand{\vect}{\boldsymbol}

\def\barroman#1{\sbox0{#1}\dimen0=\dimexpr\wd0+1pt\relax
  \makebox[\dimen0]{\rlap{\vrule width\dimen0 height 0.06ex depth 0.06ex}%
    \rlap{\vrule width\dimen0 height\dimexpr\ht0+0.03ex\relax 
            depth\dimexpr-\ht0+0.09ex\relax}%
    \kern.5pt#1\kern.5pt}}
    
\renewcommand{\eqref}[1]{\ref{#1}}

\renewcommand{\figref}[1]{Fig.\nolinebreak[4]\hspace{0.25em}\nolinebreak[4]\ref{#1}}
\renewcommand{\theequation}{\textbf{\arabic{equation}}}

\begin{document}

\title{Receptor arrays optimized for natural odor statistics}

\date{January 12, 2016}

\author{David Zwicker}
\email{dzwicker@seas.harvard.edu}
\homepage{http://www.david-zwicker.de}
\affiliation{School of Engineering and Applied Sciences, Harvard University, Cambridge, MA 02138, USA}
\affiliation{Kavli Institute for Bionano Science and Technology, Harvard University, Cambridge, MA 02138, USA}

\author{Arvind Murugan}
\email{amurugan@uchicago.edu}
\affiliation{Department of Physics and the James Franck Institute, University of Chicago, Chicago, IL 60637, USA}

\author{Michael P. Brenner}
\email{brenner@seas.harvard.edu}
\affiliation{School of Engineering and Applied Sciences, Harvard University, Cambridge, MA 02138, USA}
\affiliation{Kavli Institute for Bionano Science and Technology, Harvard University, Cambridge, MA 02138, USA}

\begin{abstract}
Natural odors typically consist of many molecules at different concentrations. It is unclear how the numerous odorant molecules and their possible mixtures are discriminated by relatively few olfactory receptors. Using an information-theoretic model, we show that a receptor array is optimal for this task if it achieves two possibly conflicting goals: (i) each receptor should respond to half of all odors and (ii) the response of different receptors should be uncorrelated when averaged over odors presented with natural statistics. We use these design principles to predict statistics of the affinities between receptors and odorant molecules for a broad class of odor statistics. We also show that optimal receptor arrays can be tuned to either resolve concentrations well or distinguish mixtures reliably. Finally, we use our results to predict properties of experimentally measured receptor arrays. Our work can thus be used to better understand natural olfaction and it also suggests ways to improve artificial sensor arrays.

\end{abstract}


\maketitle

\newlength{\figwidth}
\setlength{\figwidth}{86mm}
\setlength{\figwidth}{\columnwidth}

\begin{figure}
	\centerline{
		\includegraphics[width=\figwidth]{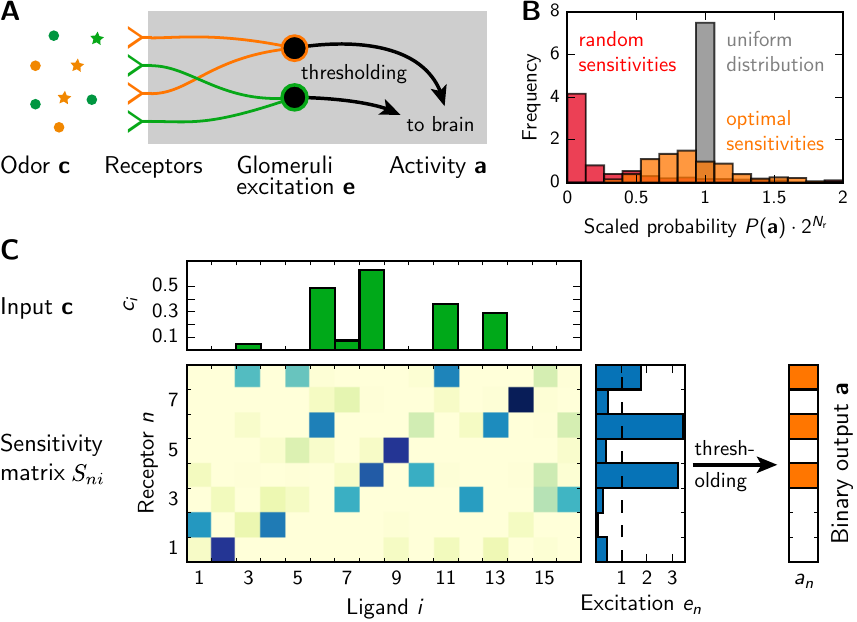}
	}
	\caption{%
	(A) Schematic representation of the olfactory system, where ligands bind to receptors, whose excitation is accumulated in glomeruli, thresholded, and relayed to the brain. 
	(B) Histogram of the probabilities~$P(\vect a)$ of the $2^{\Nr}$ output patterns~$\vect a$ for a random receptor array~(red, entropy $I=\unit[6.15]{bits}$), a numerically optimized one~(orange, $I=\unit[7.83]{bits}$), and the theoretical optimum of a uniform distribution~(green, $I=\unit[8]{bits}$).
	(C) Schematic representation of our physical model, where the input~$\vect c$ (green bars) is mapped to excitations (blue bars), which are turned into the output~$\vect a$ (orange) by thresholding.
	Parameters in (B) and (C) are $\Nr = 8$, $\Nl=16$, $p_i=\frac14$, and $\mu_i = \sigma_i = 1$.
	\label{fig:schematic}
	}
\end{figure}

Discrimination of olfactory signals occurs in a high-dimensional space of odor stimuli in which a large number of distinct molecules and their mixtures can be distinguished by a much smaller number of receptors~\cite{Touhara2009, Su2009, Mainland2014}.
For example, humans have about $300$ distinct olfactory receptors~\cite{Verbeurgt2014}, which can sense at least $2100$ odorant molecules~\cite{Dunkel2009} and the real number might be much larger~\cite{Touhara2009}.
Moreover, humans can differentiate between mixtures of up to 30 odorants~\cite{Weiss2012}.
Such remarkable molecular discrimination is thought to use a combinatorial code~\cite{Hopfield1999, Malnic1999}, where typical odorant molecules bind to receptors of multiple types~\cite{Touhara2009, Mainland2014}.
Each receptor type is expressed in many cells~\cite{Hasin2008} and the information from all receptors of the same type is accumulated in corresponding glomeruli in the olfactory bulb~\cite{Maresh2008}, see \figref{fig:schematic}A.
The activity of a single glomerulus is thus the total signal of the associated receptor type, so the information about the odor is encoded in the activity pattern of the glomeruli~\cite{Leon2003}.
This activity pattern is interpreted by the brain to learn about the composition and the concentration of the inhaled odor.
We here study how receptor arrays can maximize the transmitted information.

It is known \cite{Laughlin1981, Ruderman1994} that the input-output characteristics of sensory apparatuses of many organisms are tailored to the statistics of the organism's natural environment to maximize information transmission.
For example, in the visual circuit of the fly, the input-output relationship of neurons is matched to the cumulative distribution of the input distribution~\cite{Laughlin1981}.
Similar observations have since been made in many sensory systems \cite{Lewicki2002, Ruderman1994} and even in transcriptional regulation \cite{Tkacik2008}.
In all these cases the distinguishable outputs of the sensory system must be dedicated to equal parts of the input distribution, which is known as Laughlin's principle~\cite{Laughlin1981} or histogram equalization~\cite{Hummel1977}.
Intuitively, more of the response range is dedicated to common stimuli, at the expense of less frequent stimuli \cite{Laughlin1981}. 

Similarly, the binding affinities of olfactory receptors might reflect the natural statistics of odors in an organism's environment.
Odors vary across environments and differ in both their frequency and composition~\cite{Wright2005}.
For example, some molecules might frequently appear together because they originate from the same source while others are rarely found in the same odor.
Additionally, some odors are more important to recognize than others, which corresponds to considering an increased frequency for these odors.
Together, the frequencies and correlations constitute the natural \textit{olfactory scene}.

It is not clear how olfactory receptors can account for natural odor statistics.
Merely dedicating more receptors to common odors is not optimal, given the small number of available receptors and the many-to-many relationship between receptors and odors.
Further, the value of a receptor is strongly dependent on how it complements the other receptors in the array; many `good' receptors can still create a poor array. 
Finally, the concentrations of molecules composing an odor can vary widely.
Odors need to be distinguished both in quality and quantity;
hence receptors must vary both in what molecules they respond to and how strongly they do this.
Given the statistics of an olfactory scene, what combination of odorants should different receptors in an array respond to?

We use an information-theoretic approach to quantify how well a receptor array is matched to given odor statistics.
We generalize Laughlin's principle to the high-dimensional case and show that optimal receptor arrays should obey two general principles:

\begin{enumerate}
	\item Each receptor should be active half the time when odors are presented with natural statistics.
	\item The activities of any pair of receptors should be uncorrelated when averaged over all odors presented with natural statistics.
\end{enumerate}

If both conditions are satisfied for an array of $\Nr$ receptors with binary readouts, all $2^\Nr$ activity patterns are equally likely when odors are presented with natural statistics, see \figref{fig:schematic}B. 
However, these conditions are usually not simultaneously satisfiable.
We thus also determine the relative costs of violating the two conditions and use this to carry out numerical and analytical optimizations to determine conditions for optimal receptor arrays.

After introducing our general framework below, we first discuss general properties of optimal receptor arrays.
We then consider two different classes of natural statistics, for which we find optimal receptors in terms of random matrices.
Here, our information-theoretic approach provides a combined measure of the array's performance in multiple aspects -- from the resolution of ligand concentrations to the discrimination of mixture composition.
We thus finally discuss the trade-off between such potentially mutually exclusive goals and compare our results to experimentally measured receptor arrays.

\section{Results}
\label{sec:results}

Odors are mixtures of odorant molecules that are ligands of olfactory receptors.
Any odor can be described by a vector~$\vect c = (c_1, c_2, \ldots, c_{\Nl})$ that specifies the concentrations of all $\Nl$ possible ligands. During a single sniff, the ligands in the odor~$\vect c$ come in contact with the~$\Nr$ different odor receptors.
In the simplest case, the sensitivity of receptor~$n$ to ligand~$i$ can be described by a single number~$S_{ni}$ and the total excitation~$e_n$ of receptor~$n$ is given by~\cite{McGann2005, Lin2006}
\begin{align}
	e_n &= \sum_i S_{ni} c_i
	\;.
	\label{eqn:excitation}
\end{align}

Typical receptors have a non-linear dose-response curve~\cite{Reisert2009} and the output~$a_n$ is thus a non-linear function of~$e_n$.
Moreover, receptors are subject to noise~\cite{Lowe1995}, \eg, from stochastic binding, which limits the number of distinguishable outputs.
To capture both effects, we consider receptors with only two output states, which corresponds to large noise~\cite{Koulakov2007}.
In this case, the activity~$a_n$ of receptor~$n$ is given by
\begin{align}
	a_n &= \begin{cases}
		0 & e_n < 1 \\
		1 & e_n \ge 1	
	\end{cases}
	\;,
	\label{eqn:receptor_activity}
\end{align}
\ie, the receptor is active if its excitation~$e_n$ exceeds a threshold.
\Eqsref{eqn:excitation}--\eqref{eqn:receptor_activity} describe the mapping of the odor~$\vect c$ to the activity pattern~$\vect a=(a_1, a_2, \ldots, a_{\Nr})$, where the receptor array is characterized by the sensitivity matrix~$S_{ni}$, see \figref{fig:schematic}C.
This activity pattern is then analyzed by the brain to infer the odor~$\vect c$.
Such a distributed representation of odors in activity patterns has been compared to compressed sensing \cite{Stevens2015};
here we focus on how this representation can be tuned to match the structure of natural odors. 

We assume that the structure of natural odors in a given environment can be captured by a probability distribution $P_{\rm env}(\vect c)$ from which odors are drawn.
$P_{\rm env}(\vect c)$ can encode, for example, the fact that some ligands are more common than others or that some ligands are strongly correlated or anti-correlated in their occurrence. 
Since natural odor statistics are hard to measure~\cite{Wright2005},  we work with a broad class of distributions~$P_{\rm env}(\vect c)$ characterized by a few parameters.
We define $p_i$ to be the probability with which ligand~$i$ occurs in a random odor.
The correlations between the occurrence of ligands are captured by a covariance matrix~$p_{ij}$.
We expect $p_i$ to be small since any given natural odor typically contain tens to hundreds of ligands~\cite{Knudsen1993, Lin2006}, which is a small subset of all $\Nl \gtrsim 2100$ ligands~\cite{Wright2005}.
When a ligand~$i$ is present, we assume its concentration $c_i$ has mean~$\mu_i$ and standard deviation~$\sigma_i$.
Thus, the full natural odor statistics $P_{\rm env}(\vect c)$ are parameterized by $p_i$, $\mu_i$, and $\sigma_i$ for all ligands~$i$ and a covariance matrix~$p_{ij}$ in our model.

\subsection{Optimal receptor arrays}
An optimal receptor array must tailor receptor sensitivities $S_{ni}$ so that the odors-to-activity mapping 
given by \Eqsref{eqn:excitation}--\eqref{eqn:receptor_activity}
dedicates more activity patterns to more frequent or more important odors as specified by $P_{\rm env}(\vect c)$. In information-theoretic terms, the array must maximize the mutual information $\MI(\vect c, \vect a)$~\cite{Atick1992}. 
In our model, the mapping from~$\vect c$ to~$\vect a$ is deterministic and $\MI$ can be written as the entropy of the output distribution~$P(\vect a)$, 
\begin{align}
	\MI &= -\sum_{\vect a} P(\vect a) \log_2 P(\vect a)
	\;,
	\label{eqn:MI_def}
\end{align}
where the sum is over all possible activity patterns~$\vect a$.
Note that $P(\vect a) = \int \diff \vect c \, P(\vect a | \vect c) P_{\rm env} (\vect c)$, where $P(\vect a | \vect c)$ describes the mapping from $\vect c$ to $\vect a$.
Consequently, $\MI$ depends on $S_{ni}$ and the odor environment $P_{\rm env}(\vect c)$.
In fact, $\MI$ is maximized by sensitivities~$S_{ni}$ that are tailored to $P_{\rm env}(\vect c)$ such that all activity patterns~$\vect{a}$ are equally likely \cite{Atick1992,Laughlin1981}.

The mutual information~$\MI$ can be approximated~\cite{Sessak2009} in terms of the mean activities $\mean{a_n}$ and the covariance between receptors, $\cov(a_n, a_m)= \mean{a_n a_m} - \mean{a_n} \mean{a_m}$, encoded by $P(\vect a)$,
\begin{align}
	I &\approx 
	- \!\sum_n \bigl[\mean{a_n}\log_2 \mean{a_n} + (1 - \mean{a_n})\log_2(1 - \mean{a_n})\bigr]
	\notag \\ & \quad
	- \frac{8}{\ln 2} \sum_{n<m} \! \cov(a_n, a_m)^2
	\label{eqn:MI_est}
	\;,
\end{align}
which is an expansion up to quadratic order in $\cov(a_n, a_m)$.
The first term gives the information gained through each receptor in isolation.
The second term describes the reduction of information due to correlations between different receptors.
For both \Eqsref{eqn:MI_def} and \eqref{eqn:MI_est}, the maximal mutual information of $\unit[\Nr]{bits}$ can only be obtained if
\begin{align}
	\mean{a_n}^* &= \frac12
&
	\cov(a_n, a_m)^* &= 0
	\;.
	\label{eqn:optimality_goals}
\end{align}
Consequently, in a receptor array optimized for its natural environment, each 
receptor responds to about half of all odors 
and any  pair of receptors is uncorrelated in its response to odors, assuming odors are presented with frequency $P_{\rm 
env}(\vect c)$.

These design principles follow from very general considerations, but they may 
not always be simultaneously achievable.
To understand such constraints, we study how microscopic properties of receptor 
arrays (the sensitivities $S_{ni}$) determine both $\mean{a_n}$ 
and $\cov(a_n, a_m)$.
The mean receptor activity $\mean{a_n}$ is given by the probability that the 
associated excitation~$e_n$ exceeds~$1$, \mbox{$\mean{a_n} = 1 - F_n(1)$},
where $F_n(e_n)$ denotes the cumulative distribution function of~$e_n$, see SI.
The covariance $\cov(a_n, a_m)$ can be estimated in terms of $	\cov_c(e_n, 
e_m) $ using a normal approximation around the maximum of~$I$, see SI.
These statistics of~$e_n$ can be calculated from~\Eqref{eqn:excitation} and read
\begin{subequations}
\label{eqn:con_en_stats}
\begin{align}
	\mean{e_n}_c &= \sum_i S_{ni} \mean{c_i}
	\label{eqn:con_en_mean}
\\
	\cov_c(e_n, e_m) &= \sum_{i,j} S_{ni}S_{mj} \cov(c_i, c_j)
	\label{eqn:con_en_cov}
	\;,
\end{align}
\end{subequations}
where $\mean{c_i}$ and $\cov(c_i, c_j)$ follow from~$P_{\rm env}(\vect c)$.
Combining \Eqsref{eqn:MI_est} and \eqref{eqn:con_en_stats} to estimate mutual 
information, we can quantify how well an array's sensitivites 
$S_{ni}$ are matched to natural odor statistics $P_{\rm env}(\vect c)$. As a 
computational matter, these equations also allow a rapid calculation of mutual 
information without calculating the full distribution~$P(\vect a)$.

\subsection{Random sensitivity matrices}
We next study which sensitivity matrices~$S_{ni}$ obey the optimization goals given in \Eqref{eqn:optimality_goals} for given odor statistics.
Here, we will show that random $S_{ni}$ with independent and identically distributed entries drawn from the right distribution can be close to optimal.
This is because such matrices generically have low correlations and the resulting activities~$a_n$ are thus only weakly correlated.
In the following, we study what distributions lead to \mbox{$\mean{a_n}=\frac12$} and under what conditions these matrices minimize $\cov(a_n, a_m)$ for two different classes of odor distributions.

\paragraph{Narrow concentration distributions}
We begin with the simple case where the concentration distributions are narrow, $\sigma_i \ll \mu_i$.
In this case, we can focus on determining which ligands appear in a mixture.
Receptors that are optimal for this task must be highly sensitiv to some ligands while they ignore the others, but the exact value of the sensitivity does not matter.
This property can be encoded in a binary sensitivity matrix~$\hat S_{ni}$ where $\hat S_{ni} = 1$ if receptor~$n$ reacts to ligand~$i$ and $\hat S_{ni}=0$ if it does not.
We can then calculate activity statistics using \Eqsref{eqn:receptor_activity} and \eqref{eqn:con_en_stats}, as 
shown in the SI.
In the simple case of uncorrelated mixtures ($p_{ij} = 0$ for $i \neq j$)  
$	\mean{a_n} \approx \sum_i \hat S_{ni} \, p_i $ and 
$\cov(a_n, a_m) \approx \sum_i \hat S_{ni} \hat S_{mi} \, p_i$. In the SI, we 
also calculate corrections due to the correlated appearance of ligands ($p_{ij} 
\neq 
0$);
\eg, \mbox{$\mean{a_n} \approx 
\mean{a_n}_0 + 
\frac12(1 - \mean{a_n}_0)\sum_{i,j}(\hat 
S_{ni} + \hat 
S_{nj} - \hat S_{ni}\hat S_{nj})p_{ij}$}, 
where $\mean{a_n}_0 = \sum_i \hat S_{ni} \, p_i$ is the receptor activity in 
the uncorrelated case.

In the case of uncorrelated mixtures, we find using \Eqref{eqn:optimality_goals} that 
$\hat S_{ni}$ for optimal receptor arrays must satisfy
\begin{align}
	\sum_i \hat S_{ni}^* \, p_i  &= \frac{1}{2}
&
	\sum_i \hat S_{ni}^* \hat S_{mi}^* \, p_i &= 0
	\label{eqn:receptor_activity_binary}
	\;.
\end{align}
Receptors are thus optimal if (i) the occurrence probabilities~$p_i$ of the 
ligands they 
react to add up to~$\frac12$ and (ii) no ligand activates multiple receptors. 
Since any given ligand is rare in natural odors, $p_i \ll \frac12$, such 
optimization is equivalent to a 
partition problem where the $\Nl$ probabilities~$\set{p_i}$ have to be put into 
$\Nr$ groups (\ie, a group of ligands for each receptor), such that the sum of 
the elements is close to $\frac12$, while a 
minimal number of elements should appear in several groups. 
\Eqref{eqn:MI_est} gives the relative cost of violating these two 
possibly conflicting requirements.

\begin{figure}[t]
	\centerline{
		\includegraphics[width=\figwidth]{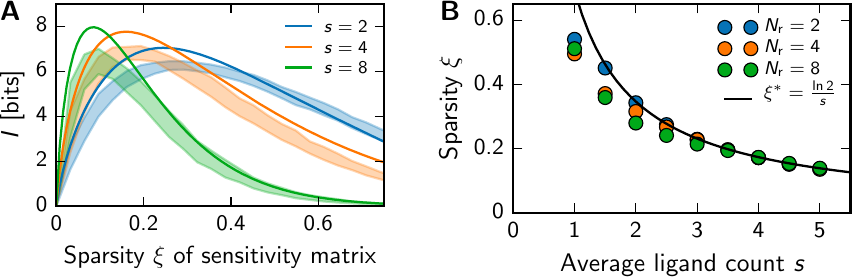}
	}
	\caption{%
	Receptor arrays with random sensitivity matrices whose sparsity $\xi$ is tuned to match natural statistics achieve near-optimal information transmission of odor composition. 
	(A) Information~$\MI$ gained by $\Nr = 8$ receptor as a function of the average sparsity $\xi$ of random binary sensitivity matrices for mixtures made of $s$ ligands drawn from a total of $\Nl = 32$  ligands.
	 Numerical results (shaded areas; mean $\pm$ standard deviation; 32 samples) and analytical results (lines) following from \Eqref{eqn:MI_est} are shown.
	(B) Sparsity $\xi$ of general binary sensitivity matrices that were numerically optimized for maximal~$I$ (symbols) is compared to the prediction from random binary matrices (solid line, \Eqref{eqn:bin_library_opt}) for different $s$ and $\Nr$ at $\Nl = 128$.
	\label{fig:bin_hom}
	}
\end{figure}%

This partition problem can be solved approximately using random binary sensitivity matrices.
The ensemble of such matrices is characterized by a single parameter, the fraction of non-zero entries or sparsity~$\xi$.
\figref{fig:bin_hom}A shows that there is an optimal sparsity~$\xi^*$, at which $\MI$ is maximized.
It follows from $\mean{a_n} = \frac12$ that
\begin{align}
	\xi^* \approx 
	\frac{\ln 2}{s}
	\label{eqn:bin_library_opt}
	\;, 
\end{align} 
where $s=\sum_i p_i$ is the mean mixture size, see SI.
This condition for random matrices agrees well with the sparsity found from numerical optimization over all binary matrices, see \figref{fig:bin_hom}B.
However, for small~$s$ the sparsity~$\xi^*$ becomes large, which leads to significant correlations $\cov(a_n,a_m)$ and thus reduced performance.
Optimal matrices thus have a sparsity that is lower then predicted by \Eqref{eqn:bin_library_opt} for small mixture sizes~$s$, see \figref{fig:bin_hom}B.

\paragraph{Wide concentration distributions}
In reality, odor concentration vary widely and receptor arrays must thus measure both odor composition and concentrations.
The concentration of a single ligand can be measured if many receptors react to it with different sensitivities~\cite{Hopfield1999}.
The receptor array is optimal for this task if all possible outputs occur with equal frequency.
This is the case if the inverse of the sensitivities follows the same distribution as the ligand concentrations~\cite{Laughlin1981}, which is known as Laughlin's principle.
However, it is not clear how this principle can be generalized for measuring the concentration of multiple ligands simultaneously.

\begin{figure}[t]
	\centerline{
		\includegraphics[width=\figwidth]{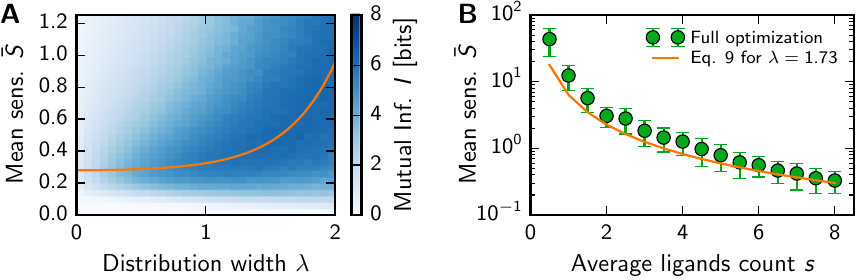}
	}
	\caption{%
		Random receptor arrays with a suitable mean sensitivity $\bar{S}$ and distribution width $\lambda$ can transmit information about both odor concentration and composition.
		(A) $\MI$ for log-normally distributed sensitivities as a function of the mean~$\bar S$ and width~$\lambda$ of the distribution for $\Nr=8$, $\Nl = 16$, $p_i = \frac{1}{4}$, and $\mu_i = \sigma_i = 1$.
		The shown mean of $\MI$ was calculated from \Eqsref{eqn:excitation}--\ref{eqn:MI_def} using Monte-Carlo sampling of 32 realizations per point.
		The orange line marks the optimum  given by \Eqref{eqn:con_dist_opt}.
		(B)~Mean sensitivity~$\bar S$ for different $s$ at $\Nr=8$, $\Nl=16$, and $\mu_i=\sigma_i=1$.
		Numerical optimizations over general sensitivity matrices (symbols; mean $\pm$ standard deviation; 64 samples) are compared to log-normally distributed matrices  (solid line, \Eqref{eqn:con_dist_opt}) with $\lambda=1.73$, equal to the mean of the numerical data.
		\label{fig:con_dist_lognorm}
	}
\end{figure}

We study this problem by considering random sensitivities that are log-normally distributed.
This choice is motivated by the complex interaction between receptors and ligands, which typically leads to normally distributed binding energies~\cite{Lancet1993}.
We will show later that experimentally measured sensitivities indeed appear to be log-normally distributed.
Log-normal distributions are characterized by two parameters, the mean~$\bar S$ and the standard deviation~$\lambda$ of the underlying normal distribution.
We thus next ask how these parameters have to be chosen to maximize the mutual information~$\MI$.
To estimate~$\MI$, we need to consider the excitations~$e_n$, which  approximately also follow a log-normal distribution~\cite{Fenton1960}.
Their statistics are given by \Eqref{eqn:con_en_stats} and read $\mean{e_n}_{c,S} = \bar S \mean{c_{\rm tot}}$ and $\cov_{c,S}(e_n, e_m) = \bar S^2 \var(c_{\rm tot}) + \delta_{nm}\var(S)\sum_i\mean{c_i^2}$, where $c_{\rm tot} = \sum_i c_i$ and $\var(S) = \bar S^2[\exp(\lambda^2) - 1]$.
We use this to calculate~$\mean{a_n}$ from \Eqref{eqn:receptor_activity} and find that the receptor array is optimal ($\mean{a_n} = \frac12$) if
\begin{equation}
	\bar S = 	
		\frac{1}{\mean{c_{\rm tot}}}
  		\left[ 1 
			+ \frac{\var(c_{\rm tot})}{\mean{c_{\rm tot}}^2}
			+ \frac{\sum_i\mean{c_i^2}}{\mean{c_{\rm tot}}^2}\left(e^{\lambda^2} - 1\right)
		\right]^{\frac12}
  \;,
  \label{eqn:con_dist_opt}
\end{equation}
see SI.
We test this equation by numerically calculating the mutual information~$\MI$ as a function of $\bar S$ and $\lambda$.
\figref{fig:con_dist_lognorm}A shows that \Eqref{eqn:con_dist_opt} predicts the optimal parameters of log-normally distributed sensitivities very well.
\figref{fig:con_dist_lognorm}B shows that this result also predicts the mean~$\bar S$ for numerical optimizations over general sensitivity matrices.

Log-normally distributed sensitivities perform badly if the distribution width~$\lambda$ is small, see \figref{fig:con_dist_lognorm}A.
This is expected since receptors with narrowly distributed~$S_{ni}$ respond similarly to all ligands, leading to large correlations~$\cov(a_n, a_m)$ and thus reduced performance~$\MI$.
Interestingly, for large enough $\lambda$ the correlations are so small that the exact value of $\lambda$ does not influence~$\MI$ significantly, see \figref{fig:con_dist_lognorm}A.
In fact, for very large~$\lambda$, the $S_{ni}$ are likely  very large or very small compared to $\bar S$.
When $\bar S$ is chosen according to \Eqref{eqn:con_dist_opt}, receptors can 
thus only detect whether ligands are present or not, corresponding to the 
binary sensitivities discussed above, which cannot resolve the concentration of 
the ligands.
Consequently, $\lambda$ must influence how well such receptor arrays can resolve concentrations.

\paragraph{Trade-off between concentration resolution and mixture discriminability}
When the distribution width~$\lambda$ is large, the receptor arrays have similar performance~$\MI$, so they are equally good at the combined problem of resolving concentrations and discriminating mixtures.
However, the performance in the individual problems can vary widely.
Since in many contexts we might wish to trade off performance, say, by sacrificing some ability to discriminate mixtures in favor of a better concentration resolution, we next investigate these properties in detail.

We define the concentration resolution~$R$ as the ratio of the concentration~$c$ at which a single ligand is presented and the concentration change~$\delta c$ that is necessary to register a change, $R=c/\delta c$.
Here, we consider the simple case where $\eta$ additional receptors have to be excited to register a change in concentration.
$R$ is a function of the concentration~$c$ at which it is measured and its maximal value
\begin{align}
	R_{\rm max} &= \frac{\Nr}{\sqrt{2\pi} \eta\lambda}
	\label{eqn:concentration_resolution}
\end{align}
 is obtained for $c=\bar S^{-1}\exp(\frac12 \lambda^2)$, which is the inverse of the median of the sensitivity distribution, see SI.

\begin{figure}[t]
	\centerline{
		\includegraphics[width=\figwidth]{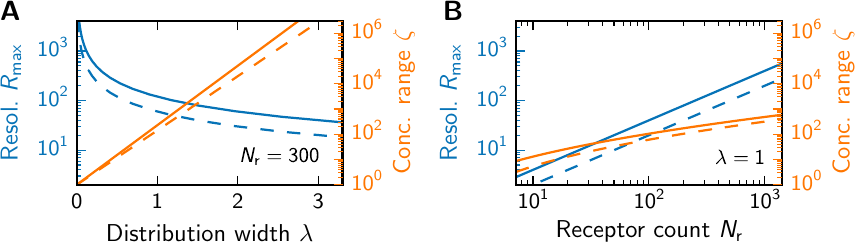}
	}
	\caption{%
		The width~$\lambda$ of the sensitivity distribution has opposing effects on concentration resolution $R_{\rm max}$ (blue, \Eqref{eqn:concentration_resolution}) and range $\zeta$ (orange, \Eqref{eqn:concentration_range}).
		(A)~$R_{\rm max}$ and $\zeta$ as a function of the width~$\lambda$ for $\Nr = 300$.
		(B)~$R_{\rm max}$ and $\zeta$ as a function of~$\Nr$ for $\lambda = 1$.
		Shown are $\eta=1$ (solid lines) and $\eta=2$ (dashed lines).
		\label{fig:con_predictions}
	}
\end{figure}%

The range of concentrations that can be detected by the receptor array is given by the ratio of  the largest concentration~$c_{\rm max}$ at which concentration differences can be detected to the lowest detectable concentration~$c_{\rm min}$, the odor detection threshold~\cite{Abraham2011}.
In terms of~$\eta$,
the logarithm of the concentration range~$\zeta = c_{\rm max}/c_{\rm min}$ reads (see SI)
\begin{align}
	\ln(\zeta) &=  2\sqrt{2} \lambda 
		\operatorname{erf}^{-1}\left(1-\frac{2\eta}{\Nr}\right) 
	\label{eqn:concentration_range}
	\;,
\end{align}
where $\operatorname{erf}^{-1}(z)$ is the inverse error function.
\Eqref{eqn:concentration_range} shows that $\lambda$ determines the number of concentration decades over which the receptor array is sensitive.

Taken together, $\lambda$ has opposing effects on the resolution and the range of concentration measurements, see \figref{fig:con_predictions}A.
Consequently, $\lambda$ can be tuned either for receptors that resolve concentrations well or cover a large concentration range.
If only single ligands are measured, the optimal $\lambda$ only depends on the concentration distribution~$P_{\rm env}(\vect c)$.
In this case, the mutual information~$\MI$ can be calculated from the resolution function~$R(c)$ and optimizing~$R(c)$ is equivalent to maximizing~$\MI$~\cite{Bialek2012}.
For odor mixtures, $\MI$ accounts for a combination of the concentration resolution and the mixture discrimination and maximizing $\MI$ does not uniquely determine an optimal receptor array.
We thus next study how the distribution width~$\lambda$ influences the ability to discriminate mixtures.

\begin{figure}[t]
	\centerline{
		\includegraphics[width=\figwidth]{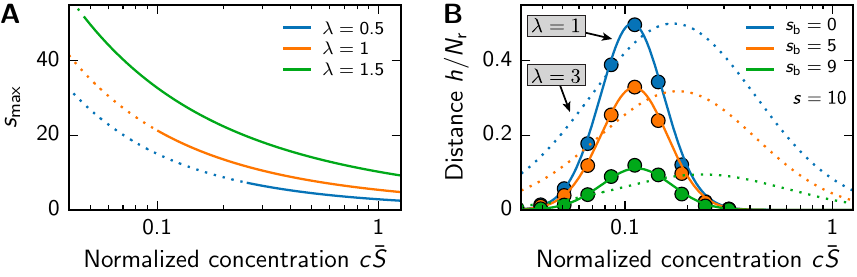}
	}
	\caption{%
		The discriminability of mixtures strongly depends on the concentrations at which odors are presented. 
		(A) Maximal mixture size~$s_{\rm max}$ (from \Eqref{eqn:mixture_size_max}) as a function of the ligand concentration~$c$ for different widths~$\lambda$ of the sensitivity distribution at $\Nr/\eta=300$.
		Dotted lines indicate where $c$ is below the detection threshold for single ligands.
		(B)~Mean difference~$h$ in the activation pattern of two mixtures of size~$s=10$ as a function of $c$ for different numbers $s_{\rm b}$ of shared ligands and widths~$\lambda$.
		Analytical results (lines) are compared to numerical simulations (symbols).
		\label{fig:mixture_discrimination}
	}
\end{figure}

We first consider mixtures of~$s$ ligands, each at concentration~$c$, and determine the maximal size~$s_{\rm max}$ where adding an additional ligand does not significantly alter the activity pattern. 
$s_{\rm max}$ is given by the largest $s$ that obeys (see SI)

\begin{align}
	\difffrac{\mean{a_n}_{S}}{s} \ge \frac{\eta}{\Nr}
	\label{eqn:mixture_size_max}
	\;,
\end{align}
where $\mean{a_n}_{S} \approx 1 - F_{\rm LN}(c^{-1}; \bar S s, \var(S)s)$ with $F_{\rm LN}(x; \mu, \sigma^2)$ being the cumulative distribution function of a log-normal distribution with mean~$\mu$ and variance~$\sigma^2$.
\figref{fig:mixture_discrimination}A shows that $s_{\rm max}$ increases with decreasing concentrations, but if the concentration falls below the odor detection threshold, individual ligands cannot be detected (dotted lines). 

Not all mixtures with less then $s_{\rm max}$ ligands can be distinguished from each other.
We show this by calculating the Hamming distance~$h$ of the activity patterns~$\vect a$ of two mixtures, \ie, the number of differences in the output.
For simplicity, we consider mixtures that contain $s$ ligands, sharing~$s_{\rm b}$ of them.
In this case, a given receptor is activated by one of the mixtures if $e_{\rm b} + e_{\rm d} > 1$, where $e_{\rm b}$ and $e_{\rm d}$ are the excitations caused by the~$s_{\rm b}$ shared and the $s - s_{\rm b}$ different ligands, respectively.
Approximating the probability distribution of the excitations as a log-normal distribution, we can calculate the expected distance~$h$, see SI.
\figref{fig:mixture_discrimination}B shows that this approximation (solid lines) agrees well with numerical calculations (symbols).
The figure also shows that mixtures can only be distinguished well if the concentration of the constituents is in the right range.
This is because receptors are barely excited for too small concentrations while they are saturated for large concentrations. 
The distance~$h$ also strongly depends on the number~$s_{\rm b}$ of shared ligands between the two mixtures, which has also been shown experimentally~\cite{Bushdid2014}.
The distance vanishes for $s_{\rm b} = s$, but \figref{fig:mixture_discrimination}B shows that  a single different ligand can be sufficient to distinguish mixtures in the right concentration range (green line).
This range increases with the width~$\lambda$ of the sensitivity distribution, similar to the range over which concentrations can be measured, see \Eqref{eqn:concentration_range}.

\subsection{Experimentally measured receptor arrays}

\begin{figure}[t]
	\centerline{
		\includegraphics[width=\figwidth]{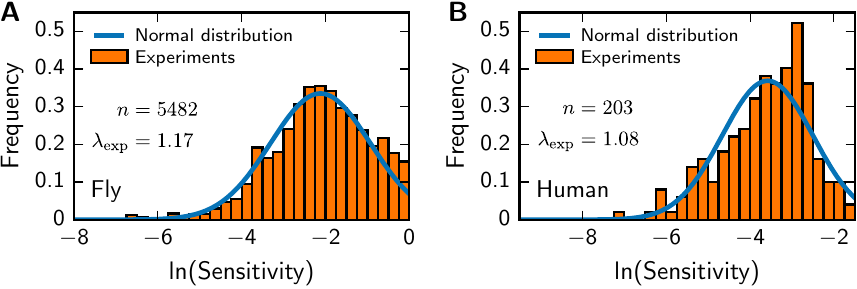}
	}
	\caption{%
		Sensitivities of olfactory receptors appear to be log-normally distributed for (A) flies~\cite{Muench2015} and (B) humans~\cite{Mainland2015}.
		The histograms of the logarithms of $n$ entries of the sensitivity matrix (orange) are compared to a normal distribution (blue) with the same mean and standard deviation~$\lambda_{\rm exp}$.
		\label{fig:sensitivities}
	}
\end{figure}

The response of receptors to individual ligands has been measured experimentally for flies~\cite{Muench2015} and humans~\cite{Mainland2015}.
We use these published data to estimate the statistics of realistic sensitivity matrices as described in the SI.
\figref{fig:sensitivities} shows the histograms of the logarithms of the sensitivities for flies and humans.
Both histograms are close to a normal distribution, with similar standard deviations~$\lambda_{\rm exp} \approx 1.1$, which implies log-normally distributed sensitivities.
Using a simple binding model between receptors and ligands, $\lambda_{\rm exp}$ can also be interpreted as the standard deviation of the interaction energies, see SI.
Consequently, these interaction energies exhibit a similar variation on the order of one $\kb T$ for both organisms, which could be caused by the biophysical similarity of the receptors.

We next use the measured log-normal distribution for the sensitivities to compare the concentration resolution~$R$ predicted by \Eqref{eqn:concentration_resolution} to measured 'just noticeable relative differences'~$R^{-1}$~\cite{Koulakov2007}.
For humans ($\Nr = 300$), the measured values are as low as \unit[4]{\%}~\cite{Cain1977}, which implies $\eta\lambda \approx 4.8$.
Using $\lambda \approx 1.1$, this suggest that about 4 receptors have to be activated until a change in concentration can be registered.
Additionally, our theory predicts that humans can sense concentrations over about $2.6$ orders of magnitude, which follows from \Eqref{eqn:concentration_range} for $\lambda = 1.1$, $\eta = 1$, and $\Nr = 300$.
However, we are not aware of any measurements of the concentration range for humans.

Our theory also predicts the maximal number of ligands that can be distinguished as a function of the concentration~$c$ of the individual ligands.
For $\lambda \approx 1.1$, we expect that the maximal number~$s_{\rm max}$ of ligands in a mixture is around 20 if individual ligands can be detected, see \figref{fig:mixture_discrimination}A.
Experimental studies report similar numbers, \eg, $s_{\rm max} \approx 15$~\cite{Jinks1999} and $s_{\rm max} < 30$~\cite{Weiss2012}.
However, \figref{fig:mixture_discrimination}A shows that $s_{\rm max}$ strongly depends on the concentration of the individual ligands and thus on experimental details.
Similarly, how well mixtures can be discriminated also depends strongly on the ligand concentration.
\figref{fig:mixture_discrimination}B shows that the concentration range over which mixtures can be distinguished is less than an order of magnitude for $\lambda \approx 1.1$.

\section{Discussion}

We studied how arrays of olfactory receptors can be used to measure odor mixtures, focusing on the combinatorial code of olfaction, \ie, how the combined response of multiple receptors can encode the composition (quality) and the concentration (quantity) of odors.
Such arrays are optimal if each receptor responds to about half of the encountered odors and the receptors have distinct ligand binding profiles to minimize correlations.

Our simple model of binary receptors can in principle distinguish a huge number of odors, since there are $\sim10^{90}$ different output combinations for $\Nr=300$.
However, it is not clear whether all outputs are achievable and how they are used to distinguish odors.
We showed that the mean receptor sensitivity must be tailored to the mean concentration to best use the large output space.
Another important parameter of receptor arrays is the fraction of receptors that is activated by a single ligand, which is equivalent to the sparsity~$\xi$ in the simple case of binary sensitivities.
If $\xi$ is small, combining different ligands typically leads to unique output patterns that allow to identify the mixtures, but the concentration of isolated ligands cannot be measured reliably, since only few receptors are involved.
Conversely, if $\xi$ is large, mixtures of multiple ligands will excite almost all receptors, such that neither the odor quality nor the odor quantity can be measured reliably.
However, here, the concentration of an isolated ligand can be measured precisely.
We discussed this property in detail for sensitivities that are log-normally distributed, where the width~$\lambda$ controls whether mixtures can be distinguished well or concentrations can be measured reliably.
Interestingly, experiments find that individual ligands at moderate concentration only excite few glomeruli~\cite{Saito2009}, but natural odors at native concentrations can excite many~\cite{Vincis2012}.
This could imply that the sensitivities are indeed adapted such that each receptor is excited about half the times for natural odors.

Our model implies that having more receptor types can improve all properties of the receptor array.
In particular, both the concentration resolution $R$ and the typical distance $h$ between mixtures are proportional to $\Nr$, a prediction that can be tested experimentally.
For instance, mice, with $\Nr \approx 1000$ receptor types, are very good at identifying a single odor in a mixture~\cite{Rokni2014}, but flies, with $\Nr = 52$~\cite{Muench2015}, should perform much worse.
However, quantitative comparisons might be difficult since the discrimination performance strongly depends on the normalized concentration~$c\bar S$ at which odors are presented.
In fact, we predict that mixtures can hardly be distinguished if the concentration of the individual ligands is changed by an order of magnitude, see \figref{fig:mixture_discrimination}B.

Our results also apply to artificial chemical sensor arrays known as 'artificial noses'~\cite{Albert2000, Stitzel2011}.
Having more sensors improves the general performance of the array, but it is also important to tune the sensitivity of individual sensors.
Here, sensors should be as diverse as possible while still responding to about half the incoming mixtures.
Unfortunately, building such chemical sensors is difficult and their binding properties are hard to control~\cite{Stitzel2011}.
If the sensitivity matrix of the sensor array is known, our theory can be used to estimate the information~$I_n$ that receptor $n$ contributes as
$
I_n \approx 
		H_{\rm b}(\mean{a_n})
		- \frac{4}{\ln 2} \sum_{m \neq n} \! \cov(a_n, a_m)^2
$
where $H_{\rm b}(p) = - p\log_2 p - (1 - p)\log_2(1 - p)$, such that $I = \sum_n I_n$, see \Eqref{eqn:MI_est}.
This can then be used for identifying poor receptors that contribute only little information to the overall results.

Our focus on the combinatorial code of the olfactory system certainly neglects intricate details of the system.
For instance, we consider sensitivity matrices with independent entries, but biophysical constraints will cause chemically similar ligands to excite similar receptors~\cite{Malnic1999, Hallem2006}.
This is important because it makes it difficult to distinguish similar ligands~\cite{Perez2015} and it might thus be worthwhile to dedicate more receptors to such a part of chemical space.
Additionally, receptors or glomeruli might interact with each other, \eg, causing inhibition reducing the signal upon binding a ligand~\cite{Ukhanov2010}.
We can in principle discuss inhibition in our model by allowing for negative sensitivities, but more complicated features cannot be captures by the linear relationship in \Eqref{eqn:excitation}.
One important non-linearity is the dose-response curve of individual receptor neurons~\cite{Reisert2009}, which we approximate by a step function, see \Eqref{eqn:receptor_activity}.
This simplification reduces the information capacity of a single glomerulus to $\unit[1]{bit}$, while it is likely higher in reality. 
However, we expect that allowing for multiple output levels would only increase the concentration resolution and not change the discriminability of mixtures very much~\cite{Koulakov2007}.
It would be interesting to see how such an extended model can measure heterogenous mixtures with ligands at different concentrations.

\begin{acknowledgments}
We thank Michael Tikhonov and Carl Goodrich for helpful discussions and a critical reading of the manuscript.
This research was funded by
the National Science Foundation through DMR-1435964, 
the Harvard Materials Research Science and Engineering Center DMR-1420570,
the Division of Mathematical Sciences DMS-1411694, and
the German Science Foundation through ZW 222/1-1.
MPB is an investigator of the Simons Foundation. 

\end{acknowledgments}

\bibliographystyle{unsrt}

\bibliography{bibdesk}

\widetext
\clearpage

\begin{center}
\textbf{\large Supplemental Information: Receptor arrays optimized for natural odor statistics}
\end{center}

\twocolumngrid

\setcounter{equation}{0}
\setcounter{figure}{0}
\setcounter{table}{0}
\setcounter{section}{0}
\makeatletter
\renewcommand{\theequation}{S\arabic{equation}}
\renewcommand{\thefigure}{S\arabic{figure}}
\renewcommand{\thesection}{S\arabic{section}}

\section{Receptor sensitivities}

\subsection{Equilibrium binding model}

We consider a simple model where receptors~$R_n$ get activated when they bind ligands~$L_i$.
This binding is described by the chemical reaction \mbox{$R_n + L_i \rightleftharpoons R_nL_i$}, where $R_nL_i$ is the receptor-ligands complex.
The equilibrium of the reaction is characterized by a binding constant~$K_{ni}$, which reads
\begin{equation}
	K_{ni} = \exp\left(\frac{E_{ni}}{\kb T}\right)	
	\label{eqn:binding_constant}
	\;,
\end{equation}
where $E_{ni}$ is the interaction energy between receptor~$n$ and ligand~$i$. 
In equilibrium, the concentrations denoted by square brackets obey $[R_nL_i] = K_{ni} \cdot [R_n][L_i]$.
Hence,
\begin{align}
	[R_nL_i] &= \frac{c^{\rm rec}_n K_{ni} c_i}{1 + \sum_i K_{ni} c_i}
	\label{eqn:binding}
	\;,
\end{align}
where we consider the case where multiple ligands compete for the same receptor.
Here, $c^{\rm rec}_n=[R_n] + \sum_i [R_nL_i]$ denotes the fixed concentration of receptors and $c_i = [L_i]$ is the concentration of free ligands.
We consider a simple receptor model in which the excitation~$e^{\rm rec}_n$ of a receptors of type~$n$ is proportional to the concentration of bound ligands,
\begin{align}
	e^{\rm rec}_n &= \alpha_n \sum_i  [R_nL_i]
	\label{eqn:receptor_occupancy}
	\;,
\end{align}
where $\alpha_n$ characterizes the excitability of receptor type~$n$.
As discussed in the main text, the excitations of all receptors of a given type are accumulated in the respective glomeruli, whose excitation~$e^{\rm glo}_n$ is thus given by $e^{\rm glo}_n = N^{\rm rec}_n e^{\rm rec}_n$, where $N^{\rm rec}_n$ is the number of receptors of type~$n$.
In the simple case of binary outputs, a glomerulus becomes active if its excitation exceeds a threshold~$t_n$, $a_n = \Theta(e^{\rm glo}_n - t_n)$, where $\Theta(z)$ denotes the Heaviside step function.
We consider the case $\alpha_n c^{\rm rec}_n \gg t_n$, where the glomerulus signals before the associated receptors become saturated.
In this case, we can linearize \Eqref{eqn:binding} and introduce the rescaled quantities
\begin{align}
	e_n &= \frac{e^{\rm glo}_n}{t_n}
& \text{and} &&
	S_{ni} &= \frac{\alpha_n N^{\rm rec}_n c^{\rm rec}_n}{t_n} \, K_{ni}
	\label{eqn:sensitivities}
\end{align}
to obtain \Eqsref{eqn:excitation}--\eqref{eqn:receptor_activity} of the main text.

A simple theory~\cite{Lancet1993} predicts that the interaction energies~$E_{ni}$ between receptors and ligands are normal distributed.
For the receptor model described above, this implies log-normal distributed binding constant~$K_{ni}$, see \Eqref{eqn:binding_constant}.
In this case, the sensitivities~$S_{ni}$ will also be log-normal distributed, see \Eqref{eqn:sensitivities}.

\subsection{Measured receptor sensitivities}
Response matrices have been measured experimentally for flies~\cite{Muench2015} and humans~\cite{Mainland2015}.
The fly database has been constructed by merging data from many studies that used various methods to measure receptor responses~\cite{Muench2015}.
It contains a non-zero response for 5482 receptor-ligand pairs, covering all 52 receptors that are present in flies.
\figref{fig:sensitivities}A in the main text shows the histogram of the logarithm of the associated sensitivities together with a normal distribution with the same mean and variance as the data.

The only comprehensive study of human olfactory receptors used a luciferase assay to measure receptor responses \textit{in vitro}~\cite{Mainland2015}.
They report the intensity of clones of 511 human olfactory receptors in response to various concentrations of 73 ligands.
Typically, the intensity of a given receptor-ligand pair is monotonously increasing as a function of ligand concentration~$c$.
We normalize the intensity to lie between $0$ and $1$ and fit a hyperbolic tangent function to determine the concentration~$c_*$ at which the normalized intensity reaches $0.5$.
Here, the only fit parameters are the concentration~$c_*$  and the slope of the tangent function at this point.
We exclude poor fits, where the relative error in either parameters is above $\unit[50]{\%}$.
This leaves us with 203 of the 623 receptor-ligand combinations, for which we then define the sensitivity as $c_*^{-1}$.
\figref{fig:sensitivities}B in the main text shows the histogram of the logarithm of these sensitivities together with a normal distribution with the same mean and variance as the data.

\section{Receptor response}

We next discuss the statistics of receptor response as a function of the odor statistics~$P_{\rm env}(\vect c)$.
We first analyze binary mixtures, where ligands are either present or not, and then consider the more complex case of continuous mixtures, which require a distribution of sensitivities to be able to sense different concentrations.

\subsection{Binary mixtures}
We consider statistics of binary mixtures ($c_i \in \set{0,1}$) that are given by
\begin{align}
	P_{\rm env}(\vect c) &= \frac{1}{Z_J[h]} \, \exp\biggl(\sum_{i,j} J_{ij} c_i c_j + \sum_i h_i c_i\biggr)
	\label{eqn:Pc_correlation}
	\;,
\end{align}
where $h_i$ denotes the commonness of ligand~$i$ and~$J_{ij}$ parameterizes correlations between ligands~$i$ and $j$.
Without loss of generality, $J_{ij}$ is symmetric with zeros on the diagonal.
The associated partition function~$Z_J$, which ensures that  $\int \! \diff \vect c \, P_{\rm env}(\vect c) = 1$, reads
\begin{align}
	Z_J[h] &= \int \! \diff \vect c \, e^{J_{ij} c_i c_j + h_i c_i}
	\;,
\end{align}
where the integral is over all binary mixtures. 
Note that we here and below use the Einstein summation convention, \ie we imply summation over repeated indices in a formula.

\paragraph{Uncorrelated binary mixtures}
For uncorrelated mixtures ($J_{ij} = 0$), the partition function reads $Z_0[h] = \int \! \diff \vect c \, e^{h_ic_i}$.
The probability~$p^*_i = \mean{c_i}_h$ of finding a ligand then reads
\begin{align}
	p^*_i
		= \frac{1}{Z_0[h]}\int \! \diff \vect c \, c_i \, e^{h_j c_j}
		= \frac{e^{h_i}}{1 + e^{h_i}}
	\;,
\end{align}
where the notation $\mean{\cdot}_h$ and the index * denote the average with respect to uncorrelated mixtures.
The covariance $p^*_{ij} = \mean{c_ic_j}_h - \mean{c_i}_h\mean{c_j}_h$ follows from
\begin{align}
 	 \mean{c_ic_j}_h
		&= \frac{1}{Z_0[h]}\int \! \diff \vect c \, c_i c_j \, e^{h_k c_k}
	= p^*_i p^*_j - \delta_{ij} (p^*_i)^2 + \delta_{ij} p^*_i
	\label{eqn:cicj_uncorr}
\end{align}
and reads 
\begin{align}
	p^*_{ij} &= \delta_{ij} p^*_i(1 - p^*_i)	
	\;.
\end{align}
The receptor activity~$a_n$, given by \Eqref{eqn:receptor_activity} in the main text, is a function of the excitation~$e_n = S_{ni} c_i$.
For binary mixtures, the step-function in \Eqref{eqn:receptor_activity} can be approximated by 
\begin{align}
	a_n \approx 1 - e^{-\gamma e_n}
	\label{eqn:gain_approx}
	\;,
\end{align}
which becomes exact in the limit $\gamma\rightarrow\infty$.
We use this to calculate the moments of $\bar a_n = 1 - a_n$,
\begin{align}
	\mean{\bar a_n}_h &= \frac{Z_0[h^{(n)}]}{Z_0[h]}
& 
	\mean{\bar a_n \bar a_m}_h &= \frac{Z_0[h^{(nm)}]}{Z_0[h]}
	\;,
\end{align}
where
\begin{subequations}
\begin{align}
	h^{(n)}_i &= h_i - \gamma \Sbin_{ni}
\\
	h^{(nm)}_i &= h_i - \gamma (\Sbin_{ni} + \Sbin_{mi})
	\;.
\end{align}
\end{subequations}
In particular, we have in the limit $\gamma\rightarrow\infty$,
\begin{subequations}
\begin{align}
	Z_0[h] &= \prod_i \left(1 + e^{h_i }\right)
\\
	Z_0[h^{(n)}] &= 
      \prod_i \left[1 + e^{h_i}(1 - \Sbin_{ni})\right]
\\
	Z_0[h^{(nm)}] &=
      \prod_i \left[1 + e^{h_i}(1 - \Sbin_{ni})(1 - \Sbin_{mi})\right]
      \;.
\end{align}
\end{subequations}

Hence,
\begin{subequations}
\label{eqn:receptor_activity_binary_exact}
\begin{align}
	\mean{a_n}_h &= 1 - \prod_i (1 - \Sbin_{ni} p^*_i)
	\label{eqn:receptor_activity_binary_mean_exact}
\\
	\cov_h(a_n, a_m) &= \prod_i 
   \bigl[ 1 - (\Sbin_{ni} + \Sbin_{mi} - \Sbin_{ni}\Sbin_{mi}) p^*_i \bigr]
\notag\\&\quad
  - \prod_i 
		(1 - \Sbin_{ni}p^*_i)(1 - \Sbin_{mi}p^*_i)
	\;.
\end{align}
\end{subequations}
We develop these equations to linear order in~$p^*_i$ to obtain \Eqsref{eqn:receptor_activity_binary} of the main text.

The receptor activity for binary sensitivity matrices with independent and identically distributed entries is described by
\begin{subequations}
\begin{align}
	\mean{a_n}_h & = 1 - \prod_i(1 -\xi p^*_i) \approx \xi s,
\\
	\cov_h(a_n, a_m)  & \approx \xi^2 s
	\;,
	\label{eqn:bin_qn_qnm_avgc_avgS}
\end{align}
\end{subequations}
where $s = \sum_i \mean{c_i}_h$ is the mean number of ligands in a mixture.
Here, $\xi$ denotes the sparsity of $\Sbin_{ni}$, which is the only parameter of the random ensemble.
The optimal sparsity~$\xi^*$ at which $\MI$ is maximized is given by the condition~$\mean{a_n}_h = \frac12$.
Using \Eqref{eqn:receptor_activity_binary_mean_exact} and solving for~$\xi$, we obtain
\begin{align}
	\xi^* \approx \Nl \frac{1 - 2^{-\frac1\Nl}}{s}
	\label{eqn:bin_library_opt}
	\;, 
\end{align}
which for large $\Nl$ at constant~$s$ becomes $\xi^* = s^{-1}\ln 2$.

\paragraph{Correlated binary mixtures}
We consider weakly correlated mixtures, where we expand all results to linear order in~$J_{ij}$.
Hence,
\begin{align}
	Z_J[h] &\approx Z_0[h] \cdot \left(1 +
			J_{ij} \mean{c_i }_h \mean{c_j}_h
		\right)
	\;.
\end{align}
The probability \mbox{$p_i = \mean{c_i}_J$} with which ligand~$i$ appears reads
\begin{align}
	p_i &= \frac{1}{Z_J[h]} \int \! \diff \vect c \, c_i \, e^{J_{jk} c_j c_k + h_j c_j}
\notag\\[2pt]
	&\approx \mean{c_i}_h \bigl[1 - J_{jk} \mean{c_j}_h \mean{c_k}_h \bigr]
	 + J_{jk} \mean{c_i c_j c_k}_h 
	\;,
\end{align}
where
\begin{align}
	\mean{c_ic_jc_k}_h &= 
		p^*_i p^*_j p^*_k
		+ \delta _{ij} \bar p^*_i p^*_i p^*_k 
		+ \delta _{ik} p^*_i \bar p^*_i p^*_j
\notag\\ &\quad
		+ \delta _{jk} p^*_i p^*_j \bar p^*_j
		+ \delta _{ij}\delta_{jk} p^*_i \bar p^*_i \left(1 - 2p^*_i\right)
\end{align}
with $\bar p_i^* = 1 - p_i^*$.
Hence,
\begin{align}
	p_i &\approx p^*_i \bigl[1 + 2J_{ij} \bar p^*_i  p^*_j \bigr]
	\; ,
	\label{eqn:pi_correlated}
\end{align}
where we used $J_{ij} = J_{ji}$ and $\operatorname{diag}(J) = \vect 0$.
Similarly, the covariance~$p_{ij} = \mean{c_ic_j}_J - \mean{c_i}_J \mean{c_j}_J$ between ligands can be calculated from~$\mean{c_ic_j}_J$, which involves
\begin{multline}
	J_{kl}\mean{c_ic_jc_kc_l}
		= 
		\delta_{ij} p^*_i \bar p^*_i  \bigl[
			2J_{ki} p^*_k (1 - 2p^*_i)
			+J_{kl} p^*_l p^*_k 
		\bigr]
\\
	+ p^*_i p^*_j \bigl[
		2 J_{ij} \bar p^*_i \bar p^*_j
		+ 2 J_{il} \bar  p^*_i p^*_l 
		+ 2 J_{jl} \bar p^*_j p^*_l 
		+ J_{kl} p^*_l p^*_k
	\bigr]
\end{multline}
and thus reads
\begin{align}
	\mean{c_ic_j}_J  &\approx
		\mean{c_i c_j}_h \bigl[1 - J_{kl} \mean{c_k}_h \mean{c_l}_h \bigr]
		+ J_{kl}\mean{c_i c_j c_k c_l}_h
\notag\\
	&=
	 p^*_i p^*_j \bigl[
		1
		+ 2J_{ij} \bar p^*_i \bar p^*_j
		+ 2J_{il} \bar p^*_i p^*_l 
		+ 2J_{jl} \bar p^*_j p^*_l
	\bigr]
\notag\\&\quad
	+ \delta_{ij} p^*_i \bar p^*_i  \bigl[
		1 
		+ 2J_{il} p^*_l\left(1 - 2p^*_i \right) 
	\bigr]
	\;,
	\label{eqn:pij_correlated}
\end{align}
where we used \Eqref{eqn:cicj_uncorr}. 
Hence,
\begin{align}
	p_{ij} &= \mean{c_ic_j}_J - \mean{c_i}_J\mean{c_j}_J
	\approx
	  \delta_{ij} p_i \bar p_i + 2J_{ij}  p^*_{ii} p^*_{jj}
	\;,
\end{align}
where $\bar p_i = 1 - p_i$.
The statistics of the receptor activity~$a_n = 1 - \bar a_n$ follow from
\begin{align}
	\mean{\bar a_n}_J &= \frac{Z_J[h^{(n)}]}{Z_J[h]}
&
	\mean{\bar a_n \bar a_m}_J &= \frac{Z_J[h^{(nm)}]}{Z_J[h]}
\end{align}
and read
\begin{subequations}
\begin{align}
	\mean{\bar a_n}_J &\approx \mean{\bar a_n}_h \cdot
		\frac{1 + J_{ij} p^{(n)}_i p^{(n)}_j}
			{1 + J_{ij} p^*_i p^*_j}
\\[4pt]
	\mean{\bar a_n \bar a_m}_J &\approx \mean{\bar a_n \bar a_m}_h \cdot
		\frac{1 + J_{ij} p^{(nm)}_i p^{(nm)}_j}
			{1 + J_{ij} p^*_i p^*_j}
	\;,
\end{align}
\end{subequations}
where
\begin{subequations}
\label{eqn:pn_pnm}
\begin{align}
	p^{(n)}_i &\equiv \mean{c_i}_{h^{(n)}} = p^*_i (1 - \Sbin_{ni})
\\
	p^{(nm)}_i &\equiv \mean{c_i}_{h^{(nm)}} = p^*_i (1 - \Sbin_{ni})(1 - \Sbin_{mi})
	\;.
\end{align}
\end{subequations}
Expanding the fractions, we obtain
\begin{subequations}
\begin{align}
	\mean{\bar a_n}_J &\approx \mean{\bar a_n}_h
		\bigl(1 + J_{ij} p^{(n)}_i p^{(n)}_j - J_{ij} p^*_i p^*_j\bigr)
\\[4pt]
	\mean{\bar a_n \bar a_m}_J &\approx \mean{\bar a_n \bar a_m}_h
		\bigl(1 + J_{ij} p^{(nm)}_i p^{(nm)}_j -  J_{ij} p^*_i p^*_j\bigr)
	\;.
\end{align}
\end{subequations}
Substituting \Eqsref{eqn:pn_pnm}, this becomes
\begin{align}
	\mean{a_n}_J &\approx \mean{a_n}_h + (1 - \mean{\bar a_n}_h)
		\bigl(\Sbin_{ni} + \Sbin_{nj} - \Sbin_{ni}\Sbin_{nj} \bigr) J_{ij} p^*_i p^*_j
\notag\\ &=
	\mean{a_n}_h + (1 - \mean{a_n}_h)
		\bigl(\Sbin_{ni} + \Sbin_{nj} - \Sbin_{ni}\Sbin_{nj} \bigr) 
		\frac{p_{ij}}{2\bar p^*_i \bar p^*_j}
	\;,
\end{align}
where in the last expression the $\bar p^*_i$ can also be replaced by~$\bar p_i$ to first order in $J_{ij}$.
For the simple case of a random, binary sensitivity matrix with sparsity~$\xi$, we obtain
\begin{align}
	\mean{a_n}_J &\approx
		\mean{a_n}_h
		+ p_{ij}\bigl(1 - \mean{a_n}_h\bigr)\left(\xi - \frac{\xi^2}{2}\right)
	\;.
\end{align}
In the case where the correlations are predominately positive ($p_{ij} > 0$), the frequency of individual ligands and the receptor response are increased, $p_i > p^*_i$ and $\mean{a_n}_J > \mean{a_n}_h$, respectively.
Consequently, the optimal sparsity must be smaller than in the uncorrelated case to have $\mean{a_n}_J = \frac12$.

\subsection{Continous mixtures}
We next consider mixtures where the concentrations of the individual ligands are drawn from a continuous distribution.
For simplicity, we consider uncorrelated mixtures ($J_{ij} = 0$, $p_{ij}=0$ for $i \neq j$), which are characterized by the probabilities~$p_i$.
In the case where receptors are excited by many ligands, the dot product $e_n = S_{ni} c_i$ can be approximated by another log-normal distribution~\cite{Fenton1960}, which is parameterized by the mean and variance given in \Eqsref{eqn:con_en_stats}.
The survival function of the log-normal distribution then implies
\begin{align}
	\mean{a_n} &\approx \frac{1}{2} \erfc\left[
		\frac{
			\ln \left(\frac{\sqrt{\mean{e_n}_c^2+\var_c(e_n)}}{\mean{e_n}_c^2}\right)
		}{
			\sqrt{2\ln \left(\frac{\var_c(e_n)}{\mean{e_n}_c^2}+1\right)}
		}\right]
	\label{eqn:con_q_n}
	\;.
\end{align}
Since $a_n^2 = a_n$, the associated variance reads
\begin{align}
	\var(a_n) &= \mean{a_n}(1 - \mean{a_n})
	\;,
\end{align}
which also determines the diagonal values of the covariance matrix~$\cov(a_n, a_m)$.
For $n\neq m$, we have
\begin{align}
	\mean{a_na_m} &= \int_1^{\infty}\int_1^{\infty} P_{\rm e}(e_n, e_m) \, \diff e_n \diff e_m
	\;,
\end{align}
where $P_{\rm e}(e_n, e_m)$ is the multivariate distribution of the two excitations $e_n$ and $e_m$.
We approximate~$P_{\rm e}(e_n, e_m)$ by a normal distribution, which describes the excitations $e_n$ and $e_m$ in the vicinity of $\mean{a_n} = \mean{a_m} = \frac12$.
This distribution is characterized by the means~$\mean{e_n}$ together with the covariances $\cov(e_n, e_m)$, which comprise five parameters in total.
Hence,
\begin{align}
	\mean{a_na_m} &\approx
	\frac{1}{4} 
	+\frac{1}{\sqrt{8 \pi }}\left(
		  \frac{\mean{e_n} - 1}{\sqrt{\var(e_n)}}
	  + \frac{\mean{e_m} - 1}{\sqrt{\var(e_m)}}
	\right)
\notag\\ &\quad
	+\frac{(\mean{e_n} - 1)(\mean{e_m} - 1) + \cov(e_n, e_m)}
		{2 \pi \sqrt{\var(e_n)\var(e_m)}}
	\label{eqn:spr_r_nm}
\end{align}
for $n\neq m$.
The associated covariance~$\cov(a_n, a_m)$ follows from the definition $\cov(a_n, a_m) = \mean{a_na_m} - \mean{a_n}\mean{a_m}$, where we obtain the mean~$\mean{a_n}$ by expanding \Eqref{eqn:con_q_n} around the optimal point \mbox{$\mean{e_n} = 1$} for small~$\var(e_n)$,
\begin{align}
	\mean{a_n} &\approx \frac12 + \frac{\mean{e_n} - 1}{\sqrt{2\pi \var(e_n)}}
	\label{eqn:spr_q_n_approx}
	\;,
\end{align}
which is the same approximation that also led to \Eqref{eqn:spr_r_nm}.
Consequently, we have
\begin{align}
	\cov(a_n, a_m)
	&\approx \begin{cases}
		\dfrac14 - \dfrac{(\mean{e_n} - 1)^2}{2\pi \var(e_n)} & n = m \\[10pt]
		\dfrac{\cov(e_n, e_m)}{2 \pi \sqrt{\var(e_n)\var(e_m)}} & n \neq m
	\end{cases}
	\label{eqn:spr_q_nm}
	\;.
\end{align}
The conditions for optimal sensitivity matrices, $\mean{a_n} = \frac12$ and $\cov(a_n, a_m) = 0$, can thus be expressed as
\begin{subequations}
\begin{align}
	\mean{e_n}^4 &= \mean{e_n}^2 + \var(e_n)
\\
	\cov(e_n, e_m) &= 0
	\label{eqn:spr_optimality_cond}
	\;,
\end{align}
\end{subequations}
see \Eqsref{eqn:con_q_n} and \eqref{eqn:spr_q_nm}.
For small $\var(e_n)$, this reduces to $\mean{e_n} \approx 1$, which indeed leads to $\mean{a_n} = \frac12$ in the approximation given in \Eqref{eqn:spr_q_n_approx}.

\subsection{Numerical simulations}
We use a  simple two-step procedure to draw odors~$\vect c$ from the statistics~$P_{\rm env}(\vect c)$. 
First, we determine the ligands that appear in a given mixture by drawing a random binary vector~$\vect b=(b_1, b_2, \ldots, b_{\Nl})$ with $b_i \in \set{0,1}$ from
\begin{align}
	P_{\rm cor}(\vect b) &= \frac{1}{Z} \exp\bigl(
			J_{ij} b_i b_j
			+  h_i b_i
	\bigr)
	\label{eqn:Pc_correlation_bn}
	\;,
\end{align}
analogously to \Eqref{eqn:Pc_correlation}.
Here, $h_i$ and $J_{ij}$ determine $p_i$ and $p_{ij}$ according to \Eqref{eqn:pi_correlated} and \Eqref{eqn:pij_correlated}, respectively.
If ligand~$i$ appears in a mixture, \ie if $b_i=1$, its concentration~$c_i$ is drawn from a log-normal distribution with mean $\mu_i$ and standard deviation~$\sigma_i$.

Given this odor statistics~$P_{\rm env}(\vect c)$ and a sensitivity matrix~$S_{ni}$, the mutual information~$\MI$ can in principle be calculated from \Eqsref{eqn:excitation}--\eqref{eqn:MI_def} of the main text.
Calculating $P(\vect a)$ to evaluate \Eqref{eqn:MI_def} involves an integral over~$P_{\rm env}(\vect c)$ over the non-linear function given in \Eqref{eqn:receptor_activity}.
We approximate this integral using Monte Carlo sampling of the odor statistics~$P_{\rm env}(\vect c)$.
Because of the stochastic nature of Monte Carlo sampling, the calculated~$\MI$ is not exact.
Consequently, we use the stochastic, derivative-free numerical optimization method CMA-ES~\cite{Hansen2006} to optimize the sensitivity matrix~$S_{ni}$ with respect to~$\MI$ to produce \figref{fig:con_dist_lognorm}B of the main text.

\section{Properties of arrays with random sensitivities}
We study properties of receptors arrays characterized by random sensitivity matrices~$S_{ni}$ whose entries are independent and identically distributed.
Here, we consider a log-normal distribution for the sensitivities, whose probability density function~$f_{\rm S}(S)$ and cumulative distribution function~$F_{\rm S}(S)$ read
\begin{subequations}
\begin{align}
	f_{\rm S}(S) &=	
	\frac{1}{\sqrt{2 \pi } S \lambda }
	\exp\left[-\frac{1}{2 \lambda ^2}\left(\ln \frac{S}{\bar S} + \frac{\lambda ^2}{2}\right)^{\!2\,}\right]
\\[8pt]
	F_{\rm S}(S) &=
	\frac12 \erfc\left[
		-\frac{1}{\sqrt{2} \lambda}\left(\ln\frac{S}{\bar S} + \frac{\lambda^2}{2} \right)
	\right]
\end{align}
\end{subequations}
and are parameterized by the mean~$\bar S$ and the width~$\lambda$, which is the standard deviation of the underlying normal distribution.
Note that all following calculations could also be performed for other sensitivity distributions.

\subsection{Concentration resolution}
\label{sec:concentration_resolution}
The fraction~$\Phi_1(c)$ of receptors that are activated by a single ligand at concentration~$c$ reads
\begin{align}
	\Phi_1(c) &= 1 - F_{\rm S}(c^{-1})
	\;.
\end{align}
The typical concentration change~$\delta c$ that is necessary to excite $\eta$ additional receptor is then defined by the condition~$\Phi_1(c + \delta c) - \Phi_1(c) = \eta \Nr^{-1}$.
Expanding~$\Phi_1$ around~$c$, the solution for~$\delta c$ reads
\begin{align}
	\delta c(c) &= \frac{\eta}{\Nr \Phi_1'(c)}
		= \frac{\eta c^2}{\Nr F_{\rm S}'(c^{-1})}
	\;.
\end{align}
For log-normal distributed sensitivities, the maximum of the associated resolution~$R=c/\delta c$ is given in \Eqref{eqn:concentration_resolution} of the main text.

\subsection{Concentration range}
The minimal concentration~$c_{\rm min}$ that can be sensed is defined by the condition~$\Phi_1(c_{\rm min}) = \eta/\Nr$, while $c_{\rm max}$ is given by $\Phi_1(c_{\rm max}) = 1 -  \eta/\Nr$.
Solving these equations,  the concentration range~$\zeta = c_{\rm max}/c_{\rm min}$ becomes
\begin{align}
	\zeta &= \frac{G_{\rm S}\bigl(1 - \frac{\eta}{\Nr}\bigr)}{G_{\rm S}\bigl(\frac{\eta}{\Nr}\bigr)}
	\;,
\end{align}
where $G_{\rm S}(y)$ is the inverse function of the cumulative distribution function $F_{\rm S}(x)$.
For log-normal distributed sensitivities, we obtain \Eqref{eqn:concentration_range} of the main text.

\subsection{Maximal number of distinguishable ligands}
\begin{figure}[t]
	\centerline{
		\includegraphics[width=\figwidth]{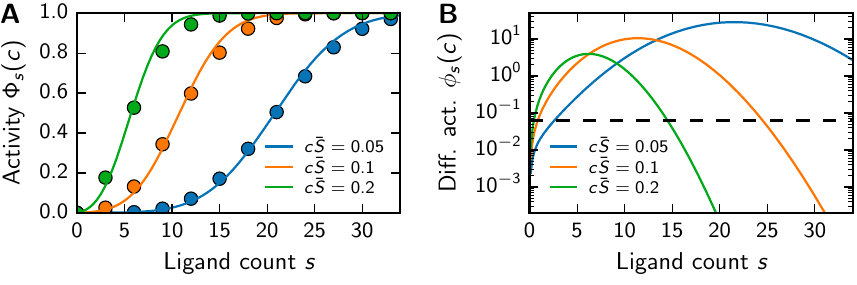}
	}
	\caption{%
		Receptors are most sensitive to mixtures of medium size.
		(A) $\Phi_s(c)$ as a function of $s$ for various~$c$ at $\lambda=1$.
		\Eqref{eqn:receptor_activity_mixture} (solid lines) is compared to numerical simulations (symbols).
		(B) $\phi_s = \diff \Phi_s/\diff s$ as a function of $s$ for various~$c$ at $\lambda=1$.
		The dashed line marks the threshold $\Nr^{-1}$ below which mixtures are not distinguishable for $\Nr = 300$.
		\label{fig:mixture_response}
	}
\end{figure}

In the simple case of a mixture with $s$ ligands, all at concentration~$c$, the fraction~$\Phi_s(c)$ of excited receptors is given by
\begin{align}
	\Phi_s(c)
	&= 1 - \hat F_{\rm S}(c^{-1}; s)
	\;,
\end{align}
where $\hat F_{\rm S}(z_n; s)$ is the cumulative probability function of the sum $z_n = \sum_{i=1}^s S_{ni}$.
If the $S_{ni}$ are log-normal distributed, the distribution for~$z_n$ can also be approximated by a log-normal distribution~\cite{Fenton1960}, which has mean $s\mean{S_{ni}}$ and variance $s\var(S_{ni})$.
In this case,
\begin{align}
	\Phi_s(c) &= 
		1- \frac{1}{2} \erfc\left[
			\frac{\ln \left(\frac{c \bar S s^2}{\sqrt{s (r+s)}}\right)}
				{\sqrt{2\ln \left(\frac{r+s}{s}\right)}}
		\right]
	\label{eqn:receptor_activity_mixture}
\end{align}
where $r = \bar S^{-2} \var(S) = \exp(\lambda^2) -1 $ is the squared coefficient of variation.
\figref{fig:mixture_response}A shows that \Eqref{eqn:receptor_activity_mixture} approximates the numerically determined $\Phi_s(c)$ well.

We next consider the maximal number of ligands that can be distinguished.
Here, we for simplicity consider the case where mixtures can be distinguished when they excite activity patterns that differ for at least $\eta$ receptors.
Since a mixture with $s$ components on average excites $\Nr \Phi_s$ receptors, this condition reads
\begin{align}
	N_{\rm r} \Phi_{s+1}(c) \ge N_{\rm r} \Phi_s(c) + \eta
	\;.
\end{align}
Expanding $\Phi_s(c)$ as a function of $s$, this condition can be approximated by
\begin{align}
	\phi_s(c) \ge \frac{\eta}{\Nr}
	\;,
\end{align}
where $\phi_s(c) = \diff \Phi_s(c)/\diff s$.
\figref{fig:mixture_response}B shows that this function has a single peak.
Mixtures with $s=0, \ldots, s_*$ ligands can thus all be distinguished from each other if
\begin{align}
	\phi_1(c) &\ge \frac\eta\Nr
& \text{and} &&
	\phi_{s^*}(c) &\ge \frac\eta\Nr
	\;.
\end{align}
Here, the first condition ensures that $c$ is above the odor detection threshold, while the second condition ensures that the two largest mixtures excite sufficiently different activity patterns.

\subsection{Discriminability of two mixtures of equal size}
We next consider how well two mixtures can be discriminated.
For simplicity, we consider two mixtures with each $s$ ligands of which $s_{\rm b}$ are shared.
We call these two mixtures plus ($+$) and minus ($-$) to distinguish them.
To determine the Hamming distance between the activation patterns, we first consider the excitations~$e_{\pm}$ of a single receptor caused by the two mixtures,
\begin{subequations}
\begin{align}
	e_{\pm} &= \sum_{i \in \mathcal C_{\rm b}} S_{ni} c + \sum_{i \in \mathcal C_{\pm}} S_{ni} c
	\;.
\end{align}
\end{subequations}
Here, $\mathcal C_{\rm b}$ denotes the set of ligands appearing in both mixtures, while $\mathcal C_{\pm}$ denote those only appearing in either of the mixtures.
Note that we only consider the case where the ligands appear with the same concentration~$c$.
The excitations can be rewritten as
\begin{align}
	e_{\pm} &= (z_{\rm b} + z_{\pm})c
\end{align}
where the $z_x$ are random variables.
Here, $z_{\rm b}$ is distributed according to $\hat F_{\rm S}(z; s_{\rm b})$, while $z_{\pm}$ are distributed according to $\hat F_{\rm S}(z; s - s_{\rm b})$.
The probability~$p_{\rm s}$ that the receptor activity is the same for both mixtures is given by
\begin{align}
	p_{\rm s} &= P(e_+ < 1 \land e_- < 1) + P(e_+ > 1 \land e_- > 1)
	\label{eqn:probability_same_pattern}
	\;.
\end{align}
The first term can be expressed as
\begin{multline}
	P(e_+ < 1 \land e_- < 1) = 
		\int_0^{\frac1c} \diff z_{\rm b}
		\hat f_{\rm S}(z_{\rm b}; s_{\rm b})
\\
	\cdot
		\int_0^{\frac1c - z_{\rm b}} \diff z_+
		\hat f_{\rm S}(z_+; s_{\rm d})
		\int_0^{\frac1c - z_{\rm b}} \diff z_-
		\hat f_{\rm S}(z_-; s_{\rm d})
	\;,
	\label{eqn:mixture_neither_excites}
\end{multline}
where $s_{\rm d} = s - s_{\rm b}$ is the number of ligands that are differ between the two mixtures.
Here, $\hat f_{\rm S}(z; s)$ denotes the probability density functions of $\hat F_{\rm S}(z; s)$.
\Eqref{eqn:mixture_neither_excites} can also be written as
\begin{multline}
	P(e_+ < 1 \land e_- < 1) 
\\
	= 
		\int_0^{\frac1c} \diff z_{\rm b}
		\hat f_{\rm S}(z_{\rm b}; s_{\rm b})
		\Bigl[
			\hat F_{\rm S}\Bigl(\frac1c - z_{\rm b}; s_{\rm d}\Bigr)
		\Bigr]^2
	\;.
\end{multline}
Similarly, the second term in \Eqref{eqn:probability_same_pattern} can be expressed as
\begin{multline}
	P(e_+ > 1 \land e_- > 1)  = 1 - \hat F_{\rm S}\Bigl(\frac1c; s_{\rm b}\Bigr)
\\
	+\int_0^{\frac1c} \diff z_{\rm b}
		\hat f_{\rm S}(z_{\rm b}; s_{\rm b})
		\Bigl[
			1 - \hat F_{\rm S}\Bigl(\frac1c - z_{\rm b}; s_{\rm d}\Bigr)
		\Bigr]^2
	\;,
\end{multline}
where the first term is the probability that the ligands appearing in both mixtures excite  the receptor alone.
The second term denotes the probability that although $z_{\rm b}$ is not large enough, both $z_+$ and $z_-$ are sufficient to bring the excitation above threshold.
The mean Hamming distance~$h= \Nr(1- p_{\rm s})$ between the activation patterns of the two mixtures then reads
\begin{align}
	h &= 2\Nr \int_0^{\frac1c} 
		\hat f_{\rm S}\Bigl(\frac1c - z; s_{\rm b}\Bigr)
		\hat F_{\rm S}(z; s_{\rm d})
		\bigl[1 - \hat F_{\rm S}(z; s_{\rm d})\bigr]
		\diff z
	\;.
	\label{eqn:mixture_distance}
\end{align}

\begin{figure*}
	\centerline{
		\includegraphics[width=2\figwidth]{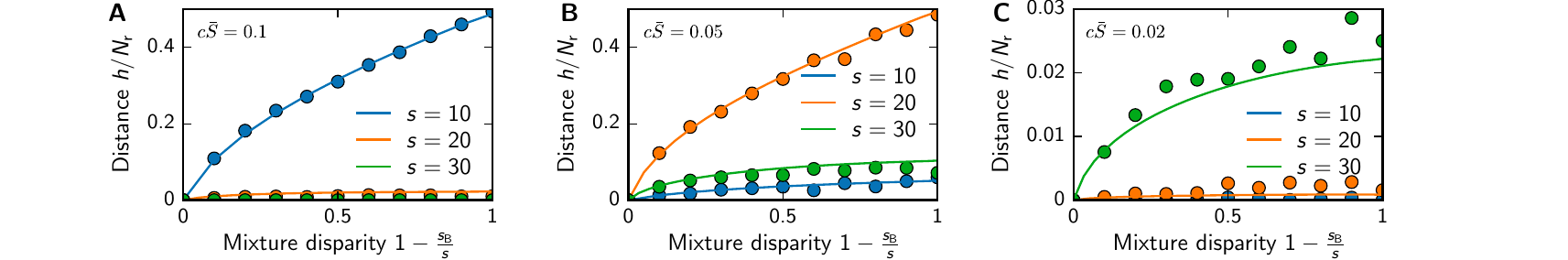}
	}
	\caption{%
		Mean normalized mixture distance $h/\Nr$ as a function of the mixture disparity $1 - s_{\rm b}/s$ for $\lambda=1$, various mixture sizes~$s$, and 
		(A)~$c\bar S = 0.1$, 
		(B)~$c\bar S = 0.05$, 
		(C)~$c\bar S = 0.02$.
		The analytical result given in \Eqref{eqn:mixture_distance} (solid lines) is compared to numerical simulations (symbols).
		\label{fig:mixture_disparity}
	}
\end{figure*}

\begin{figure}
	\centerline{
		\includegraphics[width=\figwidth]{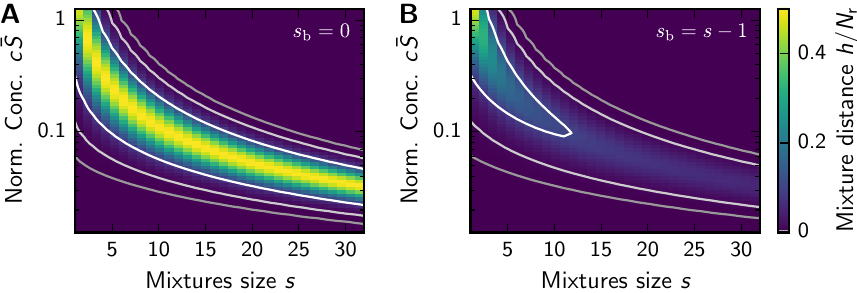}
	}
	\caption{%
		Mean normalized mixture distance $h/\Nr$ from \Eqref{eqn:mixture_distance} as a function of mixture size $s$ and concentration~$c$ of the ligands for $\lambda=1$ and
		(A)~$s_{\rm b} = 0$,
		(B)~$s_{\rm b} = s-1$.
		The lines indicate iso-contours at $h/\Nr = 0.1, 0.01, 0.001$ (white to gray).
		\label{fig:mixture_distances}
	}
\end{figure}

To test this equation, we randomly draw mixtures at given $s$ and $s_{\rm b}$, determine their activation pattern according to \Eqsref{eqn:excitation}--\eqref{eqn:receptor_activity}, and determined the associated difference.
\figref{fig:mixture_disparity} shows that \Eqref{eqn:mixture_distance} agrees well with these numerical results.
Although $h$ is a function of $s$, $s_{\rm b}$, $c$, $\bar S$, $\lambda$, and $\Nr$, the only important parameters are $s$, $s_{\rm b}$, $c\bar S$, and $\lambda$, since $\Nr$ is just a prefactor and $\bar S$ only sets the scale of typical concentrations.
We can thus explorer the behavior by plotting $h$ as a function of $s$ and $c\bar S$ for different~$s_{\rm b}$, see \figref{fig:mixture_distances}.
This plot shows that mixtures can be distinguished well when the concentration is in the right interval.
\figref{fig:mixture_distances} can be used to determine the parameter region in which a receptor array is likely able to distinguish two mixtures.
In the simple case where the activity patterns must be different in at least $\eta$ receptors, mixtures can typically be distinguished if $h>\eta$.

\end{document}